\documentclass[fleqn,usenatbib]{mnras}

\usepackage{newtxtext,newtxmath}

\usepackage[T1]{fontenc}
\usepackage{ae,aecompl}

\usepackage{graphicx}	
\usepackage{amsmath}	
\usepackage{amssymb}	

\newcommand{\vv}[1]{\mathbf{#1}}
\renewcommand{\deg}{$^\circ$~}

\title[Dawn-dusk asymmetry at comet 67P/CG]{Dawn-dusk asymmetry induced by the Parker spiral angle in the plasma dynamics around comet 67P/Churyumov-Gerasimenko}

\author[Behar, Tabone \& Nilsson]{
E. Behar$^{1,2}$\thanks{E-mail: etienne.behar@irf.se},
B. Tabone$^{3}$,
and H. Nilsson$^{1,2}$
\\
$^{1}$Swedish Institute of Space Physics, Kiruna, Sweden.\\
$^{2}$ Lule\aa\ University of Technology, Department of Computer Science, Electrical and Space Engineering, Kiruna, Sweden.\\
$^{3}$LERMA, Observatoire de Paris, PSL Research University, CNRS, Sorbonne Universit\'e, UPMC Univ. Paris 06, 75014 Paris, France
}

\date{Accepted  2018 April 26. Received 2018 April 11; in original form 2018 February 16}

\pubyear{2018}

\begin{document}
\label{firstpage}
\pagerange{\pageref{firstpage}--\pageref{lastpage}}
\maketitle

\begin{abstract}
When interacting, the solar wind and the ionised atmosphere of a comet exchange energy and momentum. Our aim is to understand the influence of the average Parker spiral configuration of the solar wind magnetic field on this interaction. We compare the theoretical expectations of an analytical generalised gyromotion with \textit{Rosetta} observations at comet 67P/Churyumov-Gerasimenko. A statistical approach allows one to overcome the lack of upstream solar wind measurement. We find that additionally to their acceleration along (for cometary pick-up ions) or against (for solar wind ions) the upstream electric field orientation and sense, the cometary pick-up ions are drifting towards the dawn side of the coma, while the solar wind ions are drifting towards the dusk side of the coma, independent of the heliocentric distance. The dynamics of the interaction is not taking place in a plane, as often assumed in previous works.
\end{abstract}




\section{Introduction}

	At comets, the sublimation of the volatiles embedded in the nucleus produces a radially expanding neutral atmosphere, which is not gravitationally bound to the body. These neutral molecules can be ionised through photoionisation, electron impact, or charge exchange, and new-born ions are added to the solar wind. From there, the new-born ions will be accelerated by the ambient electric and magnetic fields and in the absence of collisions, momentum and energy are exchanged through the fields between the solar wind and the partially ionised atmosphere. This phenomenon is known as the \textit{mass-loading} of the solar wind by the new-born ions, as mass is added to the plasma \citep{szego2000ssr}. On large scales, much larger than the scales of the ion gyromotion, the cometary ions are seen accelerated instantaneously at the average plasma velocity, a result of the ideal magnetohydrodynamics (MHD, see for example \citet{flammer1991ass, schmidt1993jgr}). However, in order to understand how momentum and energy are exchanged between the two populations, one has to consider smaller scales, and thus resolve the gyromotion of the ions\footnote{For most purposes, at comets, the hybrid approximation is relevant, and electrons can be considered as a massless and charge-neutralising fluid.} (see for example the simulation works of \citet{hansen2007ssr, rubin2014icarus, koenders2016mnras}). \textit{In-situ} results were obtained on the cometary ion gyromotion at different comets (Halley, Giacobini-Zinner, Grigg-Skjellerup and Borelly) and reviewed by \citet{coates2004asr}. The present work is based on \textit{Rosetta} data, taken in the close environment of comet 67P/Churyumov-Gerasimenko (67P/CG). 

	In the absence of gravity, the motion of charged particles in an electric field $\vv{E}$ and a magnetic field $\vv{B}$ is dictated by the Lorentz force, $\vv{F}_{\mathcal{L}} = q\ (\vv{E} + \vv{v} \times \vv{B})$, with $q$ the charge of the particle and $\vv{v}$ its velocity. In some cases this motion may be very simple. In the undisturbed solar wind for instance, the effect of the electric field and of the magnetic field are cancelling each other, and the resulting acceleration on the solar wind particles is null. When considering the simplistic case of a single new-born ion added to the solar wind with no initial velocity, one finds the classical cycloidal motion with the Larmor radius $\mathcal{R} = m\ v_\perp/(q\ |\vv{B}|)$. In this case, the solar wind and the electric- and magnetic fields are left unaffected by the addition of a single ion. If the source region of cometary ions is much larger than their gyroradius, gyrotropic (phase space) distribution functions can form. First, unstable ring distributions form, which become thickened shell distributions to eventually transform into Maxwellian distributions. The theory and the observations of these gyrotropic distributions are reviewed by \citet{coates2004asr}. The deceleration of the solar wind can then be tackled by a fluid approach.
	
	 At 67P/CG however, for heliocentric distances large enough, the cometary ion gyroradius was comparable or larger than their source region, and the cometary pick-up ion distribution functions are non-gyrotropic \citep{behar2018aa_ns, koenders2016mnras} (we note that closer to the Sun, more complex distribution functions of cometary ions were observed on two cases, and reported by \citet{nicolaou2017mnras}). It was shown that in this context, as the cometary pick-up ion density becomes comparable to the solar wind ion density, both populations gyrate, and the fluid description of the solar wind breaks down. From upstream of the coma to the close tail region, individual solar wind ions describe less than one period of their gyromotion \citep{behar2017mnras, behar2018aa_model, behar2018aa_ns}. Cometary ions are initially accelerated along the electric field, and the solar wind is accelerated (deflected) in the opposite direction, a clear result observed within the coma of 67P/CG \citep{behar2016grl, bercic2018aa}. 
In these studies, the dynamics of both populations were depicted to take place in a plane, containing the comet-Sun line and the upstream electric field. In a semi-analytical model of the solar wind dynamics proposed in \citet{behar2018aa_model}, the same hypothesis is done by assuming that the upstream magnetic field is perpendicular to the upstream solar wind velocity. However, on average, the magnetic field has an angle with the flow direction different than 90\deg, because of the Parker spiral configuration of the interplanetary magnetic field (IMF). The influence of the IMF angle on the interaction between the solar wind and different obstacles has been studied at other unmagnetised bodies, and often result in similar dawn-dusk asymmetries, as shown by the simulation work of \citet{jarvinen2013jgr} at Venus, or by \textit{in-situ} results at Mars \citep{dubinin2008pss} and at the Moon \citep{harada2015jgr}.

 In the present short article, we explore the effect of such an angle on the interaction between the solar wind and the cometary ions, with a straightforward statistical approach based on measurements at comet 67P/CG as well as IMF measurements at 1 au. Using statistics over a long period allows us to overcome the one-point-measurement limitations (\textit{i.e.} the space coverage and the time coverage are always at the great cost of one another). \\

\section{Generalised gyromotion}
	
	In \citet{behar2018aa_model}, the dynamics of the interaction between two perfect beams of plasma (the solar wind and the cometary ions, of respective velocities and densities $\vv{u}_{sw}$, $n_{sw}$ and $\vv{u}_{com}$, $n_{com}$) is analytically obtained for length scales over which the total electric field is reduced to
	\begin{equation}
	\arraycolsep=1.4pt\def\arraystretch{1.8}
	\begin{array}{rl}
		\vv{E} = & -\vv{u}_i \times \vv{B}\\
		\vv{u}_i = & \dfrac{n_{com}\vv{u}_{com} + n_{sw}\vv{u}_{sw}}{n_{com}+n_{sw}}
	\end{array}
	\label{E}		
	\end{equation}
	
	Introducing the expression of the electric field in the Lorentz force gives the following equation of motion for each beam, assuming $q_{sw}=q_{com}=q$~:
	
	\begin{equation}
	   \begin{array}{|rcl}
       		\dot{\vv u}_{sw} & = & ~~~~\dfrac{q \ n_{com}}{m_{sw}(n_{sw}+n_{com})} ( {\vv u}_{sw} - {\vv u}_{com} ) \times \vv{B} \\
	        ~ \\
	        \dot{\vv u}_{com} & = &  -\dfrac{q \ n_{sw}}{m_{com}(n_{sw}+n_{com})} ( {\vv u}_{sw} - {\vv u}_{com} ) \times \vv{B}
	   \end{array}
	   \label{dynamics}
	\end{equation}
	
\begin{figure}
   	\begin{center}
  		\includegraphics[width=.5\textwidth]{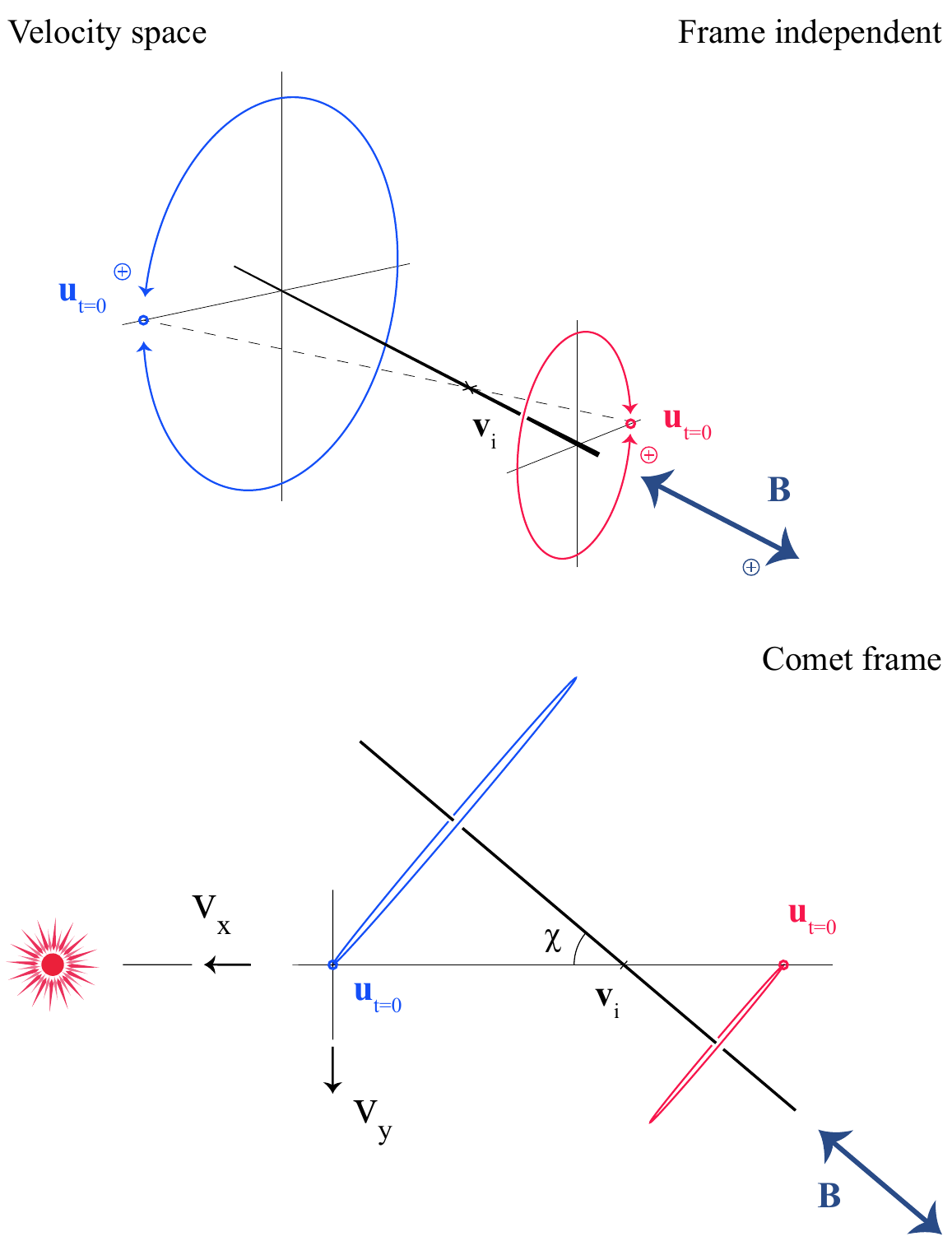}
      		\caption{Evolution in velocity space of two interacting plasma beams (generalised gyromotion), for two opposite senses of the magnetic field, frame independent (top) and within a chosen frame (bottom). In this frame, the magnetic field lies in the $(x,y)$-plane. Applied to the situation at the comet, blue is used for the cometary ions, and red for the solar wind ions.}
      		\label{genGyr}
   	\end{center}
	\end{figure}

	All ions of a same population experience the same force at the same time, and the single particle velocity is equal to the population average velocity: a beam remains a beam. Elements of the resolution of these equations are given in \citet{behar2018aa_model}. One finds that in the most general configuration of the initial conditions, the two beams will evolve in velocity space along circles, with each circle contained in a plane orthogonal to the magnetic field: there is no acceleration along the magnetic field. Therefore, if initially the beams have different velocity components parallel to $\vv{B}$, the two circles are in different planes and the motion in physical space is not happening in a plane in any frame. The top plot of Figure \ref{genGyr}, purely frame independent, shows the evolution of the beams in velocity space. If the magnetic field sense is flipped, the beams evolve along the same two circles, with the opposite rotations. As indicated by the + sign in the upper panel of Figure \ref{genGyr} and in agreement with the classical single particle motion, these rotations are prograde given the sense of $\vv{B}$. The velocity of the centre of mass of the two beams $\vv{v}_i$ is conserved through time, $\dot{\vv v}_i = 0$, as shown by the black crosses on the same plots. These dynamics generalise the classical single test-particle gyromotion to the case of two plasma beams in an electric and a magnetic field, with an arbitrary initial configuration.
	
	We now put the exact same configuration in a reference frame, shown in the bottom plot of Figure \ref{genGyr}. There, cometary ions in blue have no initial velocity, the beam is at the origin at $t=0$. At the same time, the solar wind velocity is chosen to be along the $x$-axis, which points at the Sun. The $z$-axis completes this right-handed cartesian frame, referred to as the comet frame. In this precise frame, the magnetic field has an angle $\chi$ with the $x$-axis, and is within the $(x,y)$-plane. The $v_z$-component for each species is changing sign every half gyration period. More interestingly, we find that the solar wind beam has a $v_y$-component, which is always positive. Conversely, the cometary ions have a $v_y$-component always negative. The evolution of the $v_y$-component is the same no matter the sense of the magnetic field.\\
	
	In the comet frame, one can easily derive the velocity of the guiding centres of each population (centres of the two circles), which correspond to the drift of the populations in physical space. In this frame, the cometary ions are drifting perpendicular to the magnetic field, similarly as in the illustration in \citet{coates2004asr}, Figure 2. When solar wind ions are largely dominating, this tends to the classical $\vv{E} \times \vv{B}$-drift of a test particle. As the cometary ion density gets larger, the drift speed decreases.
	
	The solar wind ions drift towards the +$y$-axis with an angle that depends on the density and mass ratios, which will therefore evolve through the coma, resulting in complex trajectories.
	
	We note however that the problem cannot be reduced to the motion of guiding centres in the case of 67P/CG. Guiding centres are not relevant since ions are only following the early phase of a single gyro-period \citep{behar2017mnras, behar2018aa_model, behar2018aa_ns}. In other words, ions do not have time to drift.
	
	\section{\textit{OMNI} dataset and \textit{Rosetta} observations}
	
	\begin{figure*}
      \begin{center}
     	 \includegraphics[width=\textwidth]{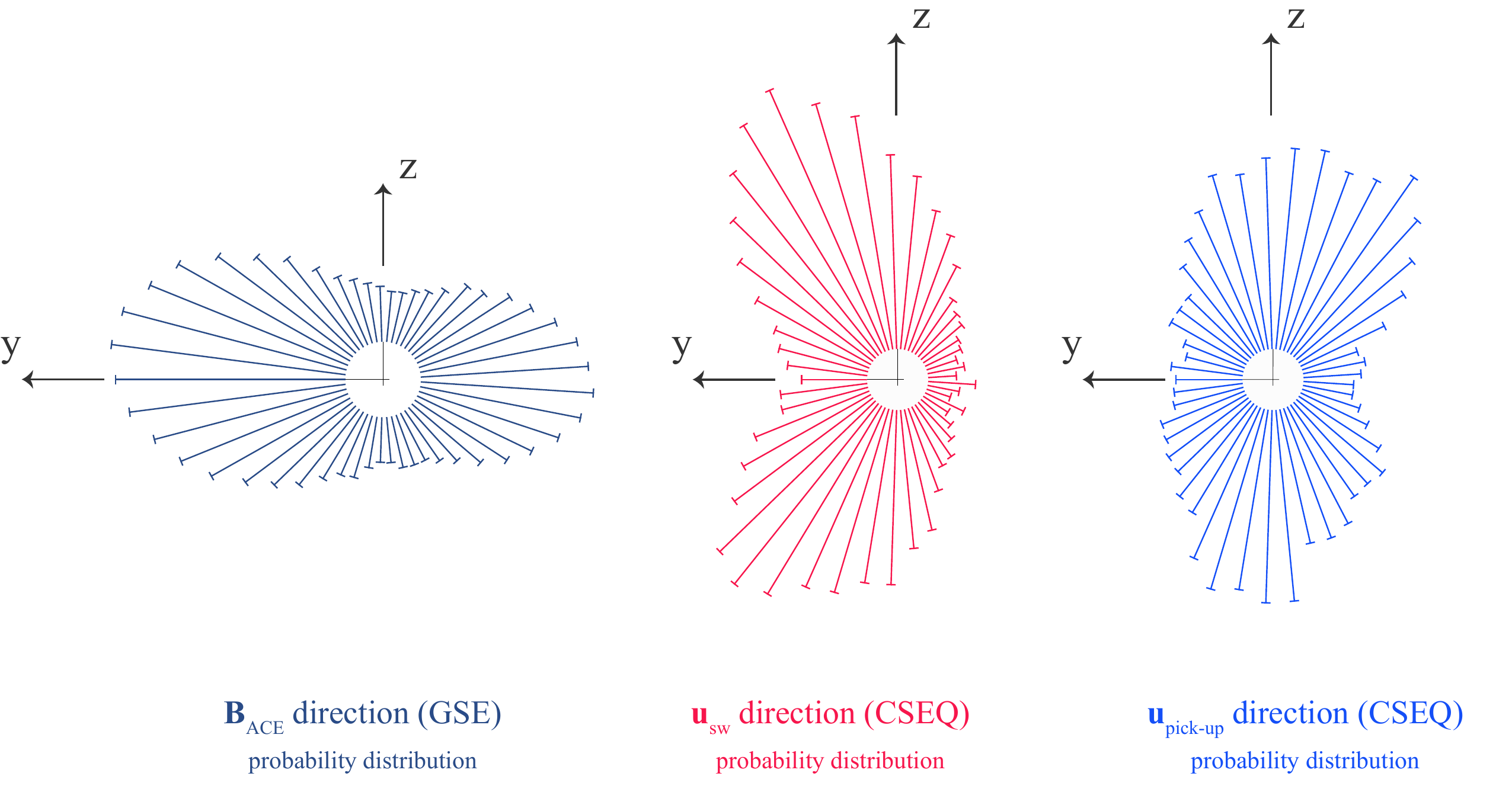}
         \caption{Probability distributions of the vector directions projected in the $(y,z)$-plane.}
         \label{distrib}
      \end{center}
   	\end{figure*}
   
	In the Sun equatorial plane, the magnetic field is on average directed along an Archimedean spiral, lying in the same plane (the Parker spiral). The local angle of the spiral is determined by the Sun's rotation, the speed of the radially expanding solar wind, and the distance to the Sun. The orbital plane of comet 67P/CG and of Earth are both inclined about 7\deg away from the Sun equatorial plane. In this study, the inclinations are neglected and we consider the average IMF upstream of the comet and upstream of the Earth to be given by the planar Parker spiral illustrated in Figure \ref{interpretation}.
	
	Within the coma of 67P/CG, the solar wind ions and the cometary ions were measured by the imaging spectrometer RPC-ICA (Rosetta Plasma Consortium - Ion Composition Analyzer, \citet{nilsson2007ssr}). The instrument has a field of view of 90\deg$\times$ 360\deg, measures positive ions with energy from about 10 eV to 40 keV, and can discriminate their masses. The duration for a complete velocity space scan is 3 minutes. The limited field of view and its obstruction by the spacecraft are not expected to induce any systematic effect, considering the constant movement of the probe, the variability of the ion dynamics, and the integration over several months of data. Using daily manual selections of the energy and mass channel range where a significant signal is seen, as presented in \citet{behar2016grl} and \citet{bercic2018aa}, one can separate solar wind protons and cometary pick-up ions throughout most of the mission. Another cometary ion population -- the new-born ions -- is observed, and is excluded from the cometary pick-up ions selection. Based on these selections, plasma moments are integrated. The aberration caused by the motion of the comet around the Sun is corrected, with barely any effect. We use here the bulk velocity of each of the two populations expressed in the Body-Centred Solar Equatorial (CSEQ) reference frame: the $x$-axis points to the Sun, the $z$-axis is perpendicular to the $x$-axis and oriented by the Sun's north pole, the $y$-axis completes the right-handed triad. The $(x,y)$-plane and the Sun equatorial plane are thus close to parallel (seperated by 7\deg, neglected here). As the rotation of the nucleus is prograde, the dusk is along the $+y$-axis, and the dawn is along the $-y$-axis. The results shown in Figure \ref{distrib} were taken between the beginning of the active mission, August 6th 2014 (3.6 au), and the end of the mission, September 30th 2016 (3.8 au). When the spacecraft is within the solar wind ion cavity (from early June 2015 to mid December 2015, see \citet{behar2017mnras}), solar wind protons are not observed, and cometary pick-up ion data are as well not considered. The resulting data set is about 21 month long.

	Solar wind magnetic field data at 1 au were retrieved from the \textit{OMNIWeb Plus} interface, have a time resolution of 5 minutes. For this precise time period, the data were obtained by the \textit{ACE} probe \citep{smith1998} and the \textit{WIND} probe \citep{lepping1995ssr}.
	The IMF direction is expressed in the Geocentric Solar Ecliptic (GSE) reference frame, with the $x$-axis pointing to the Sun, the $z$-axis orthogonal to the $x$-axis and parallel to the ecliptic north pole (7.2\deg away from the Sun rotation axis), the $y$-axis completing the right-handed system. Based on these definitions, the IMF is expected to be on average in the $(x, y)$ plane, both at Earth and 67P/CG. To that extent, the use of \textit{ACE} data is more illustrative than necessary. It also quantifies how variable the IMF was over the same period.\\
	
	The magnetic field measured by at 1 au is projected in the $(y, z)$-plane of the GSE frame, and a probability distribution function of its direction is obtained for all data measured between August 2014 and September 2016. The result is shown in Figure \ref{distrib}, left panel. In the context of this figure, the direction of a vector is given by its orientation and its sense. As expected, two peaks are found close to the $y$-axis, due to the Parker spiral average configuration of the IMF. An apparent tilt of the distribution is found, as well as an asymmetry in the maximum probability of the two peaks. The same analysis was done over a two month wide sliding window, verifying that these aspects result from statistical fluctuations. The exact same procedure is done with the proton and pick-up cometary ion bulk velocities and shown in Figure \ref{distrib} centre and right panels. All available bulk velocities (one per 3 minutes long scan) over the mission are filtered as following: only densities above $10^{-3}$ cm$^{-3}$ are considered, and only vectors that are further than 20\deg away from the Sun-comet line are considered. These two thresholds, however arbitrary, allow to consider signals in the proper range of the instrument sensitivity, and signals with a well defined orientation in the $(y,z)$-plane. Considering all data without filters on the density and the flow direction gives the exact same results, with less pronounced features. About 62000 data points are binned for the solar wind protons, and about 56000 for the cometary pick-up ions. The protons display two peaks with either a positive or a negative $z$-component, both with a positive $y$-component (dusk). Similarly, the probability distribution of the cometary ion direction has two peaks along $+z$ and $-z$, and the probability distribution has a general shift towards the $-y$ direction (dawn). This general shift appears far away from the Sun as well, at the largest heliocentric distances \textit{Rosetta} has probed, and is at 0 order constant with the heliocentric distance (not shown here).\\
\newpage
	The $x$-component of all the probability distributions of Figure \ref{distrib} depends on the heliocentric distance. The angle between the Sun-Earth line and the magnetic field orientation, measured to be of about 45\deg at 1 au, would increase and tend to 90\deg with increasing heliocentric distances. Therefore the $(x,y)$-projection of the probability distribution of the magnetic field direction measured at Earth is not relevant for the comet. Concerning the ions measured at the comet, the angle between their bulk velocity and the Sun-comet line has shown strong evolution with heliocentric distances. For instance, the solar wind deflection has been observed from about 10\deg up to 180\deg \citep{behar2017mnras}. Therefore the $(x,y)$-projection of their distribution, integrated over the entire mission, does not provide a valuable information.

	\section{Phase space distribution function and Momentum exchange}
	
		\begin{figure}
   	\begin{center}
  		\includegraphics[width=.8\columnwidth]{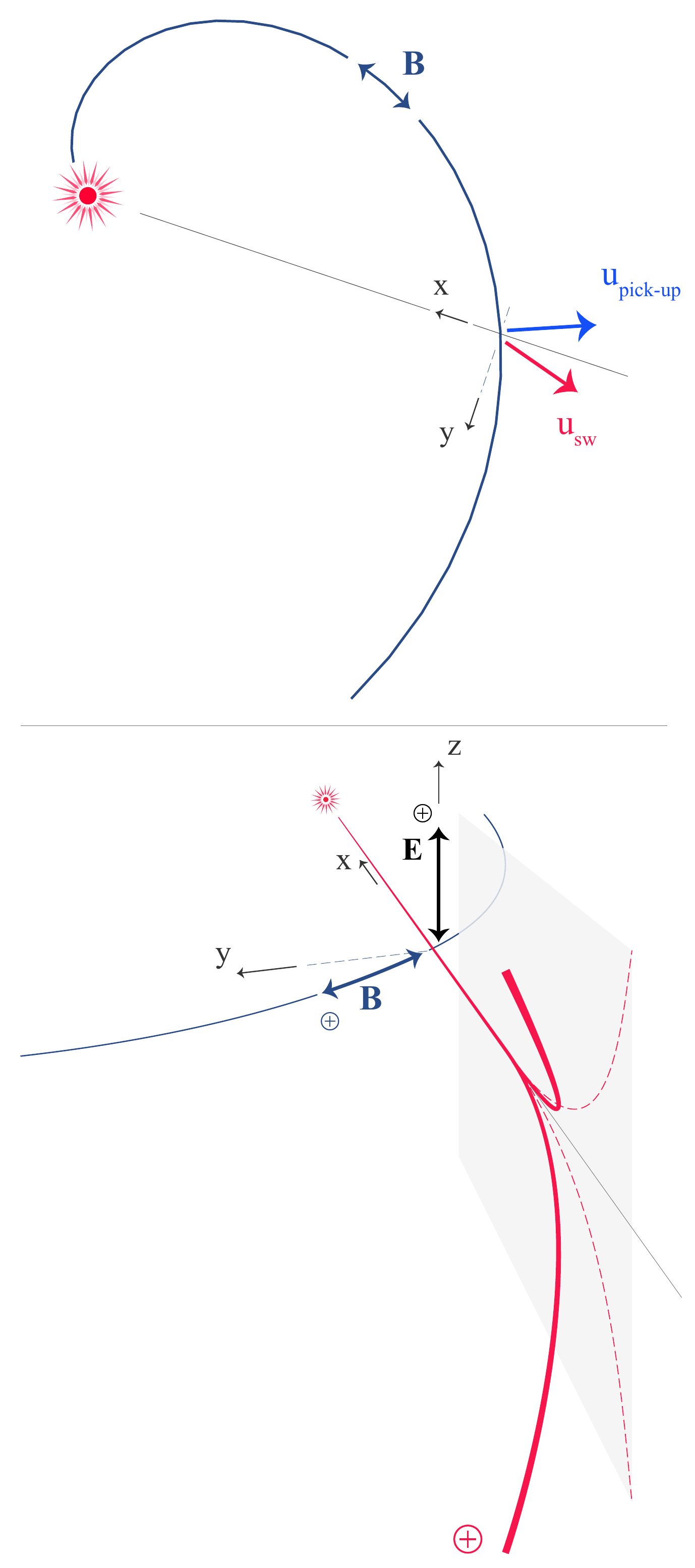}
      		\caption{Schematics of the dynamics, for the solar wind ions (red) and the cometary pick-up ions (blue). No matter the sense of the IMF along the Parker spiral, the solar wind proton velocity will gain a component along the $+y$-axis (dusk), while the cometary pick-up ions gain a negative $-y$-component (dawn).}
      		\label{interpretation}
   	\end{center}
	\end{figure}
	
	The shift towards the $-y$ direction (dawn) of the cometary pick-up ions is not as pronounced as for the solar wind protons (dusk). The reason for this may be instrumental, one possibility being that the selected pick-up ions may be contaminated by cold cometary ions (see \citet{stenberg2017mnras, bercic2018aa}). It may also be purely physical. The distribution function of cometary pick-up ions were observed to get more complex closer to the Sun, as reported by \citet{nicolaou2017mnras} in two case studies. In one of the cases, a partial ring distribution function is found for the cometary pick-up ions, compatible with their gyration in the coma. The average orientation of such a distribution function and its effects on the result of Figure \ref{distrib} is not obvious. However these partial ring distributions are still rare in ICA observations, and their average shape in the CSEQ reference frame is thus not accessible. Further statistical investigation is needed to find out if and when the cometary ion distribution function at the spacecraft location turns from non-gyrotropic to gyrotropic.
	
	We also note that additionally, the solar wind ions did not have perfect beam-like distribution functions at all time either, at the spacecraft position. In \citet{behar2017mnras}, Figure 2, partial ring distributions of solar wind protons are shown, observed at around 2 au, when the solar wind ion cavity was about to pass over the spacecraft location. 
	The fact that only partial ring distributions were observed at 67P/CG stresses out that, as discussed previously, in the day side of the coma and close to the nucleus, solar and cometary ions have only followed part of one gyration, with a gyroradius following the evolution of the density ratio between the two populations \citep{behar2018aa_model}. This is most likely why the bulk velocities exhibit such a clear statistical behaviour. 
	
	It was verified that including or not heliocentric distances below 2.5 au does not change the result, qualitatively, therefore these more complex distributions do not change the observed asymmetry at a statistical level.\\
		
	The two peaks of the probability distribution of the magnetic field direction indicate that as expected, the solar wind electric field is mostly directed along either $+z$ ($\vv{B}$ along $+y$) or $-z$ ($\vv{B}$ along $-y$, accordingly with Equation \ref{E}). Thus upstream of the measurement point, where the solar wind first meets the new-born ions, the exchange of momentum between the solar wind and the thin coma happens along the $z$-axis. However within the coma, close to the nucleus, protons are seen with a positive $y$-component and cometary ions with a negative one, consistently with the generalised gyromotion presented here-above, in the case of a non-perpendicular magnetic field. Therefore, the motion of both populations is not contained in a single plane through the coma, as illustrated in the bottom panel of Figure \ref{interpretation}. In this schematic, the exchange of momentum initially takes place in the grey plane, but immediately the proton red trajectory will get a component along the $y$-axis and get out of the plane. Depending on the direction of the magnetic field, the trajectory will be deflected towards +$z$ or -$z$, but with a +$y$-component in both cases.\\
	
	An obvious reason for the upstream magnetic field to not be perpendicular with the solar particle flow, and therefore along the $y$-axis, is the angle of its spiral configuration (Figure \ref{interpretation}).
	Despite large heliocentric distances (with a maximum distance of 3.8 au), the IMF upstream of 67P/CG always had on average an angle different than 90\deg with the $x$-axis (theoretically, the Parker angle spiral at 3.8 au is expected to be of about 75\deg \citep{cravens2004}) Thus, as shown by the generalised gyromotion and on average, no matter the sense of the IMF along the spiral and additionally to their deflection towards the $\pm z$-axis, the protons will have a positive $v_y$-component at all heliocentric distances, towards the dusk side of the coma, and the cometary pick-up ions will have a negative $v_y$-component at all heliocentric distances, towards the dawn side of the coma.\\

	\section{Conclusion}
	
	By generalising the gyromotion of two populations interacting with each other and working in a  precise reference frame, it was shown that the solar wind ions and the cometary pick-up ions are expected to drift sideway, with drifts of opposite signs, regardless of the magnetic field sense.\\
	
	The statistical approach at comet 67P/CG over a period of 21 months allowed us to overcome the lack of measurement upstream of the interaction region. Based on data, it was demonstrated that indeed both populations have a velocity component contained in the IMF plane, misaligned with the Sun-comet line, no matter the heliocentric distance. Since ions move along less than a gyro-period, this velocity component does not strictly correspond to a drift. This additional velocity component is duskward for the solar wind protons, while cometary pick-up ions have a dawnward additional velocity component, no matter if the magnetic field is outward or inward the Parker spiral.\\
	
	
	In this article we only describe an average configuration. An orientation of the IMF that would depart from the Parker spiral would obviously results in different orientations of the velocity vectors as well, which in fact correspond to the breadth of the probability distributions in Figure \ref{distrib}. Qualitatively, the configuration is only rotated around the Sun-comet line. In a plasma frame in which the upstream electric field would have the same orientation and sense at any time (the Comet Sun Electric field frame for instance, in which the electric field is along the $z$-axis, the $x$-axis pointing to the Sun), the probability distributions would be much narrower (and of different shape). However this work demonstrates that such a plasma frame cannot be obtained properly, without monitoring of the upstream solar wind parameters. In \citet{behar2017mnras} and \citet{bercic2018aa}, a proton-aligned reference frame, based on the observed solar wind proton direction, was used. We now see that such a proton-aligned frame is on average rotated from the orientation of the upstream solar wind electric field. We note that this rotation has no impact on the results of these studies.

\section*{Acknowledgements}

This work was supported  by the Swedish National Space Board (SNSB) through grants 108/12, 112/13 and 96/15. 

We are indebted to the whole {\it Rosetta} mission team, Science Ground Segment and {\it Rosetta} Mission Operation Control for their hard work making this mission possible.




\bibliographystyle{mnras}
\bibliography{cometLib} 





\bsp	
\label{lastpage}
\end{document}